\documentclass{emulateapj}
\usepackage[latin9]{inputenc}
\usepackage{array}
\usepackage{color}
\usepackage{graphicx}
\usepackage{amssymb}
\usepackage{esint}

\begin{document}

\title{Optimal Binning of the Primordial Power Spectrum}

\author{Paniez Paykari}
\email{p.paykari06@ic.ac.uk}
\author{Andrew H. Jaffe}
\email{a.jaffe@ic.ac.uk}
\affil{Imperial College London\\
Astrophysics Group, Blackett Laboratory, Imperial College, London
SW7 2AZ, UK }

\begin{abstract}
The primordial power spectrum describes the initial perturbations
in the Universe which eventually grew into the large-scale structure
we observe today, and thereby provides an indirect probe of inflation
or other structure-formation mechanisms. In this paper we will investigate
the best scales the primordial power spectrum can be probed, in accordance
with the knowledge about other cosmological parameters such as $\Omega_{b}$,
$\Omega_{c}$, $\Omega_{\Lambda}$, $h$ and $\tau$. The aim is to
find the most informative way of measuring the primordial power spectrum
at different length scales, using different types of surveys and the
information they provide for the desired cosmological parameters.
We will find the optimal binning of the primordial power spectrum
for this purpose, by making use of the Fisher matrix formalism. We
will then find a statistically orthogonal basis for a set of cosmological
parameters, mentioned above, and a set of bins of the primordial power
spectrum to investigate the correlation between the two sets. For
this purpose we make use of principal component analysis and Hermitian
square root of the Fisher matrix. The surveys used in this project
are Planck and SDSS(BRG), but the formalism can easily be extended
to any windowed measurements of the perturbation spectrum.
\end{abstract}


\section{Introduction}

\textcolor{black}{The primordial Power Spectrum (PS) probes the physics
of structure formation in the early Universe. In particular, inflation
provides a paradigm for early Universe physics in accord with current
cosmological observations. Simple models of inflation predict an almost
Gaussian distribution of adiabatic perturbations with a scale-invariant
spectrum (i.e., $P(k)\propto k$). However, there are other possibilities:
for example, there could be more than one scalar field during inflation
and this would predict a different spectrum and possibly a different
distribution of fluctuations. The details of inflation are presently
unknown to us. Therefore, determining the primordial PS would give
us better intuition about the early Universe.}

\textcolor{black}{We use different surveys --- of the Cosmic Microwave
Background, of the galaxy power spectrum, of velocity fields, etc.
--- to constrain the primordial PS. However, the spectra of these
surveys are }\textit{\textcolor{black}{jointly}}\textcolor{black}{$\;$sensitive
to cosmological parameters (which, we will collectively call $\theta_{i}$)}\textcolor{red}{
}\textcolor{black}{$\:$and the primordial PS. Hence there is a statistical
degeneracy between the two. The aim here is to explore, as the data
improve, what new information can be learnt about the primordial PS
and what exactly needs to be improved to better constrain the primordial
PS. The motivation is to test the assumptions about the initial conditions
besides getting better constraints on parameters based on the same
set of assumptions. Therefore, knowing the degeneracy between the
cosmological parameters and primordial PS, we want to investigate
the scales the primordial PS can be probed best with future experiments.}

The outcome of different surveys is usually a type of PS that is a
convolution of the primordial PS, whatever form it may have, and (the
square of) the transfer function of the particular type of survey,
which holds the cosmological parameters. Here I list some examples;

\begin{itemize}
\item For galaxy surveys, the PS is related to the primordial PS through
the matter PS, $P_{\delta}(k)$, as\begin{equation}
P_{g}(k)=b^{2}(k)P_{\delta}(k)\simeq b^{2}(k)2\pi^{2}kT^{2}(k)\Delta_{\zeta}^{2}(k)\;,\label{eq:gPS-mPS}\end{equation}
where $\Delta_{\zeta}^{2}$ is the primordial PS and $T(k)$ is the
matter transfer function and $b(k)$ is the \textit{bias}.
\item For CMB surveys, the angular PS is\textcolor{black}{\begin{equation}
C_{\ell}=4\pi\int_{0}^{\infty}d\ln k\Delta_{\ell}^{2}(k)\Delta_{\zeta}^{2}(k)\;,\end{equation}
where $\ell$ is the related to the angular scale on sky via }$\ell\sim180^{0}/\theta$\textcolor{black}{
$\;$and $\Delta_{\ell}(k)$ is the angular transfer function of the
radiation anisotropies. }
\end{itemize}
Here, we define the primoridial curvature power spectrum, which we
parameterize as $\Delta_{\zeta}^{2}(k)=A(k/0.05)^{n_{s}-1}$. $A$
is the amplitude and $n_{s}$ is the spectral index. The notation
refers to the gauge-invariant curvature perturbation $\zeta$ \citep{Bardeen-GaugeInvariant}.

\textcolor{black}{Other types of power spectra, such as the weak lensing
and peculiar velocity power spectra, have similar forms; they depend
on the cosmological parameters, through a transfer function, and the
primordial PS. These different power spectra probe different scales
with different accuracies. }One survey can, therefore, help fill the
gaps in other surveys and all together they are expected to improve
information, especially on the overlapping scales. This means combining
surveys can help us choose narrower bins and hence investigate the
primordial PS to a greater resolution.

One common method for error estimation is to use a Fisher matrix analysis.
The Fisher matrix is generally used to determine the sensitivity of
a particular survey to a set of parameters and has been largely used
for forecasting and optimisation. The Fisher matrix is the ensemble
average of the \textit{curvature} of a function $\mathcal{F}$ (i.e.
it is the average of the curvature over many realisations of signal
and noise)\begin{equation}
F_{\alpha\beta}=\left\langle \mathcal{F}\right\rangle =\left\langle -\frac{\partial^{2}\ln\mathcal{L}}{\partial\theta_{\alpha}\partial\theta_{\beta}}\right\rangle \label{eq:General_FM}\end{equation}
The Fisher matrix allows us to estimate the errors on parameters without
having to cover the whole parameter space. Hence, the inverse of the
Fisher matrix is a crude estimate of covariance matrix of the parameters,
by analogy with a Gaussian distribution in the $\theta_{a}$, for
which this would be exact. The authors of \citep{bjk} have compared
the Fisher matrix analysis with the full likelihood function analysis
and found there was great agreement between the two methods if the
likelihood function is approximately Gaussian near the peak. The Cramer-Rao
inequality states that the smallest non-marginalised error measured
for the parameters by any unbiased estimator (such as the maximum
likelihood) is $1/\sqrt{F}$
\footnote{It should be noted that the Cramer-Rao inequality is a statement about
the so-called ``Frequentist'' confidence intervals and is not strictly
applicable to ``Bayesian'' errors.
}. The marginalised%
\footnote{Integration of the joint probability over other parameters. %
} one-sigma error is $\sqrt{(F^{-1})_{\alpha\alpha}}$ for parameter
$\alpha$.

The Fisher matrix for CMB surveys is given by\begin{equation}
F_{\ell\ell^{\prime}}=f_{sky}\frac{2\ell+1}{2}\delta_{\ell\ell^{\prime}}[C_{\ell}+w^{-1}e^{\ell^{2}\sigma^{2}}]^{-2}\;,\label{eq:FM_CMB}\end{equation}
where $C_{\ell}$ is the angular PS, $w$ is the weight defined as
$(\Delta\Omega\sigma_{n}^{2})^{-1}$ with $\Delta\Omega$ being the
real space pixel size and $\sigma_{n}^{2}$ being the noise per pixel,$\;$\textcolor{black}{
$e^{-\ell^{2}\sigma^{2}}$ is the window function}\textcolor{red}{}%
\footnote{\textcolor{black}{This damps power on larger $\ell$s; as we get closer
to the resolution limit of the survey $C_{\ell}$s start to correlate.}%
}\textcolor{red}{ }$\:$for a Gaussian beam (where
$\sigma=\theta_{fwhm}/\sqrt{8\ln2}$) and $f_{sky}$
is the fraction of the sky observed. The factor
$f_{sky}(2\ell+1)$ gives the number of independent modes at
a given wavenumber; the term proportional to $C_{\ell}$ is the
sample (or cosmic) variance contribution, and the
$w^{-1}e^{\ell^{2}\sigma^{2}}$ term is the noise contribution. Note
that the diagonal form for the matrix implies diagonal
(uncorrelated) errors on the $C_{\ell}$s. To find errors on other
parameters, we use the Jacobian\begin{equation}
F_{\alpha\beta}=\sum_{\ell}F_{\ell\ell^{\prime}}\frac{\partial
C_{\ell}}{\partial\theta_{\alpha}}\frac{\partial
C_{\ell^{\prime}}}{\partial\theta_{\beta}}\;,\label{eq:CMB
FM}\end{equation} where $\theta_{\alpha}$ and $\theta_{\beta}$ are
different parameters.

For a volume-limited galaxy survey the Fisher matrix \citep{tegmark1997}$\;$is%
\footnote{\textcolor{black}{Note that this equation only applies to linear regime,
as non-linearities impose non-Gaussianities. }%
}\begin{equation}
F_{nn^{\prime}}=\delta_{nn^{\prime}}\frac{k_{n}^{2}\Delta kV}{(2\pi)^{2}(P_{n}+1/\bar{n})^{2}}\;,\label{eq:FM_g}\end{equation}
where $V$ is the total volume of the survey, \textcolor{black}{$\bar{n}$
is the number density of the survey ($N_{tot}/V$),} $P_{n}$ is the
galaxy PS in each $k_{n}$ bin and $\Delta k$ is the \textcolor{black}{binwidth.
Similar to the CMB power spectrum case, $k_{n}^{2}\Delta kV$ counts
the number of modes, $P_{n}$ gives the sample variance, and $1/\bar{n}$
the noise variance due to Poisson counting errors. This, again, gives
us the errors on the galaxy PS and we use the Jacobian to get the
errors on other parameters}\begin{equation}
F_{\alpha\beta}=\sum_{n}F_{nn^{\prime}}\frac{\partial P(k_{n})}{\partial\theta_{\alpha}}\frac{\partial P(k_{n^{\prime}})}{\partial\theta_{\beta}}\;.\label{eq:FM_gJ}\end{equation}

Fisher matrices for different surveys can easily be combined by a
simple summation $\mathbf{F}=\mathbf{F}_{\textnormal{galaxy}}+\mathbf{F}_{\textnormal{CMB}}$.
\textcolor{black}{This is because they are proportional to the $\log$
of the likelihood function and we multiply likelihoods to combine
them. Equivalently, we can think of them as the }\textit{\textcolor{black}{weights}}\textcolor{black}{
$\:$(inverse noise variance) of the experiments, which add for a
Gaussian distribution. }The nonzero correlation between the parameters
in the covariance matrix makes interpreting the errors somewhat more
difficult than the uncorrelated case. We will discuss various methods
for decorrelating the power spectra and cosmological parameters.

\section{Method}

The aim is to investigate the primordial PS in a ``non-parametric''
way (we use quotations remarks to remind the reader that ``non-parametric''
merely means that we use a very general model, potentially with a
very large number of parameters). Therefore, we assume a top-hat binning
of the primordial PS\begin{equation}
\Delta_{\zeta}^{2}(k)=\sum_{B}w_{B}(k)Q_{B}\,,\end{equation}
where $Q_{B}$ is the power in each bin $B$ and $w_{B}=1$ if $k\in B$
and $0$ otherwise. The cosmological parameters under investigation
are (and of the form) $\Omega_{c}$, $\Omega_{b}$, $\Omega_{\Lambda}$,
$h$, $\tau$ and $n_{s}$. \textcolor{black}{The reason for inclusion
of $n_{s}$ in the parameters is to allow for a consistency check.
In this setting we do not expect to see any correlation between $n_{s}$
and the bins of primordial PS. }Inclusion of $n_{s}$ in the parameter
space only makes minute changes to our results and can be ignored.
We will choose a geometrically flat (adiabatic) $\Lambda$CDM model
with WMAP5 \citep{Dunkley-WMAP} values for the parameters; $n_{s}=0.963\pm0.0145$(with
zero running), $\Omega_{m}=0.214\pm0.027$, $\Omega_{b}=0.044\pm0.003$,
$\Omega_{\Lambda}=0.742\pm0.03$\textcolor{black}{, $\tau=0.087\pm0.017$
}and $h=0.719\pm0.0265$, where $H_{0}=100$$h\textnormal{km}^{-1}\textnormal{Mpc}^{-1}$.
$\Omega_{\nu}=0.0$ was chosen, as massive neutrinos introduce some
difficulties in the Fisher matrix analysis \citep{eht} and therefore
were ignored for now. CMBfast software \citep{sz} was used for the
calculations. The surveys chosen for this initial investigation are
the projected results from the SDSS Bright Red Galaxies(BRG)\textcolor{black}{}%
\footnote{\textcolor{black}{These are bright galaxies, which means the survey
can be quite deep, with $z\sim0.25-0.5$. Also, these trace the elliptical
galaxies, which are thought to be better tracers of mass at this redshift
range. }%
} sample and the Planck Surveyor CMB Power Spectrum%
\footnote{{\small http://www.rssd.esa.int/SA/PLANCK/docs/Bluebook-ESA-SCI\%282005\%291\_V2.pdf}%
}.

\section{Galaxy Surveys --- SDSS(BRG)}

A galaxy PS is related to the matter PS via a parameter called \textit{\textcolor{black}{bias}}
--- equation {\footnotesize \ref{eq:gPS-mPS}}.\textcolor{black}{
$\;$For the BRG sample of SDSS, this is assumed linear and scale-independent
with the form $P_{g}=b^{2}P_{\delta}$, where $b$ is the bias $\:$and
approximately equal to $2.0$ \citep{mph,sw,H-AcousticOsc-SDSS-DR4,SW-LSS-bias}.
The survey specifications for BRG sample are} $\bar{n}=10^{5}/V$
and $V=(1$$h^{-1}\textnormal{Gpc})^{3}$ \textcolor{black}{\citep{gwsdss}}.

For the $\theta_{i}$ the derivatives in the Jacobian were obtained
numerically using the Taylor expansion\begin{equation}
P(\theta_{i})=P(\theta_{0})+({\normalcolor \frac{\partial P}{\partial\theta_{i}}){\color{black}{\color{red}{\color{black}\Delta(\theta_{i})}\;}.}}\end{equation}

The width and direction of the step are quite important here. A two-sided
derivative was chosen, so that the derivative is centred on the default
value $\theta_{0}$, with a step size of $\Delta(\theta_{i})/2$ on
each side. This \textcolor{black}{is} accurate to $2^{nd}$ order
in $\Delta(\theta_{i})$ (a one-sided derivative would be at a slightly
shifted place of $\theta_{i}+\Delta(\theta_{i})/2$, and is only accurate
to $1^{st}$ order \citep{eht}). The width of the step should be
small enough to give accurate results and yet big enough to avoid
numerical difficulties. This was taken to be a $5\%$ variation, therefore
a $2.5\%$ width on each side. Other studies have shown that this
turns out to be the best step size, giving the most accurate results
\citep{eht}.

\textcolor{black}{For the primordial PS bins, the derivative is proportional
to the matter transfer function}\begin{equation}
\frac{\partial P_{g}(k)}{\partial\Delta_{\zeta}^{2}(k^{\prime})}=4\times2\pi^{2}kT^{2}(k)\delta_{kk^{\prime}}\;,\label{eq:primfishmat}\end{equation}
where $k$ and $k^{\prime}$ refer to the bins. The $k$-range for
SDSS is $0.006\lesssim k/(h\textnormal{Mpc}^{-1})\lesssim0.1$. The
minimum value for the wavenumber, \textcolor{black}{$k_{min}$, is
obtained from the largest scale of the survey --- $(2\pi/V^{1/3})$.
Its maximum value, $k_{max}$, is chosen to avoid non-linearities.
Simulations of a very similar flat model \citep{mwp} suggested a
$k_{max}$ of $0.1h\textnormal{Mpc}^{-1}$. This is also very close
to the scale at which departures from linear theory was seen by \citet{PW08}.}

The derivatives in the Jacobian need to be averaged into bins. Later
we will explain the criteria for choosing the widths and locations
of the bins.

\section{CMB Surveys --- Planck}

One thing to note in this case is that the output of CMBfast is of
the form $\mathcal{C}_{\ell}=[\ell(\ell+1)/2\pi]C_{\ell}$, so the
CMB Fisher matrix, equation \ref{eq:FM_CMB}, is multiplied by this
factor. The specifications for Planck HFI ($\nu=100$GHz) are $\theta_{fwhm}=10.7^{\prime}=0.003115$
radians, $\sigma_{pix}=1.7\times10^{-6}$, $w^{-1}=0.028\times10^{-15}$
\citep{dpgl}. The derivatives in the Jacobian were again obtained
numerically by Taylor expansion\begin{equation}
\mathcal{C}_{\ell}(\theta_{i})=\mathcal{C}_{\ell}(\theta_{0})+(\frac{\partial\mathcal{C}_{\ell}}{\partial\theta_{i}})\Delta(\theta_{i})\;.\end{equation}

The same arguments as in the SDSS case applies for the width and direction
of the step here. In the case of the primordial PS bins, the derivative
becomes\footnote{To obtain $\Delta_{\ell}(k)$, CMBfast needed to be altered to give
the radiation transfer functions at all $\ell$s. Then, for each $\ell$,
this was interpolated in $k$. %
}
\begin{equation}
\frac{\partial\mathcal{C}_{\ell}}{\partial\Delta_{\zeta}^{2}(k)}=2\ell(\ell+1)\int_{k_{min}^{B}}^{k_{max}^{B}}dk\left|\Delta_{\ell}(k)\right|^{2}\;.\label{eq:Cl_pPS_derivative}\end{equation}
This needs to be averaged into $k$ bins, which will be explained
later on. \textcolor{black}{The chosen $k$-range for Planck is $0.0001\lesssim k/(h\textnormal{Mpc}^{-1})\lesssim0.1$,
where $k_{min}$ was obtained from $k_{min}=\ell_{min}/d_{A}=2/d_{A}$,
where $d_{A}$ is the angular diameter distance to the surface of
last scattering obtained to be $\sim14\textnormal{Gpc}$ \citep{Dunkley-WMAP}. }

\section{SDSS \& Planck}

\textcolor{black}{As explained above, to combine data from different
surveys, we can add the Fisher matrices obtained for each of them.
We expect to see an improvement on the errors of both the bins and
cosmological parameters. Equivalently, this will enable us to have
narrower bins without sacrificing Signal-to-Noise per bin. }

\section{Optimal Binning }

As explained before, a set of primordial PS bins are part of our parameter
space. In this section we will explain how these bins are chosen.
For our purposes the bins need to have the same amount of contribution
to the Fisher matrix which means they need to have the same $S/N$.
We take the signal in each bin to be the amplitude of the primordial
PS in that bin and the noise to be given by the inverse of the square
root of the diagonal elements of the Fisher matrix. For this, we construct
a signal vector, $\mathbf{S}$, which contains the amplitude of the
primordial PS for all the bins and the values of the cosmological
parameters. We weight our Fisher matrix by this vector\begin{equation}
F_{\alpha\beta}^{\prime}=S_{\alpha}F_{\alpha\beta}S_{\beta}\,,\label{eq:(S/N)^2 FM}\end{equation}
where there is no Einstein summation. This now gives us a $(S/N)^{2}$
matrix, where the square root of its diagonal elements are the $S/N$
for the bins, and the weighted errors for $\theta_{i}$s. It is worth
to emphasise that it is this $(S/N)^{2}$ Fisher matrix that will
be diagonalised later on.

For the SDSS case, we start by having the maximum number of bins possible
in our $k$-range. The usual properties of the Fourier transform imply
that the scale of the survey not only determines $k_{min}$, but also
puts a limit on our resolution: \textcolor{black}{$k_{min}=(\Delta k)_{min}=(2\pi/V^{1/3})$};
narrower bins would become highly correlated. Therefore, we set up
a series of bins with this minimum binwidth in our $k$-range. We
then construct a Fisher matrix for this set of bins (and $\theta_{i}$s)
and weight it by the signal vector, $\mathbf{S}$, for this set. With
this binning adopted, the $S/N$ values range from $3.7$ in the first
bin to $35.1$ in the last bin. Knowing that the binwidths chosen
are the minimum possible and that increasing binwidths will increase
the $S/N$ value, we conclude that the bin with the maximum $S/N$
cannot be changed and hence we make other bins wide enough to reach
the $S/N$ in this bin. To obtain this optimal binning we start an
iteration; \textcolor{black}{smaller bins of size $\simeq(\Delta k)_{min}/6$
are combined until the $S/N$ are equal to 15\% of the maximum $S/N$:\begin{equation}
\frac{\textnormal{Max}(S/N)-(S/N)_{i}}{\textnormal{Max}(S/N)}=0.15\,,\end{equation}
where $i$ refers to the bins. This gives us $8$ bins with their
$S/N$ ranging $30-35$.}

For Planck, the bins are obtained so that their $S/N$ matches that
of SDSS. The reason for applying this criteria to Planck is to allow
for a fair comparison between the results from SDSS and Planck. This
criteria gives us a total of $23$ bins for Planck.

In the case of the combined Planck and SDSS we require only that the
S/N of the bins are equal to 50\%.\textcolor{black}{ This now gives
us the optimal resolution of the primordial PS we can achieve from
SDSS and Planck. We have a total of $48$ bins with }$S/N$\textcolor{black}{$\:$being}
in the vicinity of $\sim20.0$ and, therefore, still comparable to
the $S/N$ values in the other cases.

It is worth reminding the reader that an aternative way to determine
the binning would be to take the marginalised errors as the noise.
This would be obtained by inverting the Fisher matrix in each iteration
loop to get the covariance matrix, which gives the marginalised variances
of the bins and $\theta_{i}$s. We would then take the sub-block of
this covariance matrix that refers to the bins only, and invert it
to get a \textit{marginalised} Fisher matrix for the bins. We would
then feed this Fisher matrix into equation \ref{eq:(S/N)^2 FM}. However,
this method could not be implemented because the SDSS Fisher matrix
is not invertible; the SDSS Fisher matrix is not a positive definite
matrix because it is asked to estimate too many parameters. There
are a total of $n$ data points ($n$ galaxy PS bins) and we are asking
these to predict $n+m$ parameters ($n$ primordial PS bins and $m$
$\theta_{i}$s). Also, note that whichever of the methods presented
uses the \textit{correlated} errors as the noise. We now discuss the
decorrelation of the parameters.

\section{Decorrelating the Parameters}

\subsection{Principal Component Analysis}

One popular method to overcome the correlation between the parameters
is to perform principal component analysis (PCA); the covariance matrix
is a symmetric $n\times n$ matrix and therefore, can be diagonalised
using its eigenvectors. This has the form $\mathbf{C}=\mathbf{E^{T}}\mathbf{\Lambda}\mathbf{E}$,$\;$where
$\mathbf{C}$ is the covariance matrix, $\mathbf{{\normalcolor E}}$
is an orthogonal matrix with the eigenvectors of ${\normalcolor \mathbf{C}}$
as its rows and ${\normalcolor \mathbf{\Lambda}}$ is the diagonal
matrix with the eigenvalues of ${\normalcolor \mathbf{C}}$ as its
diagonal elements%
\footnote{It is common to construct the covariance matrix for the PCA. However,
Fisher matrix can be used instead; eigenvectors stay the same, but
eigenvalues are inversed.%
}. This constructs a new set of variables $\mathbf{{\normalcolor X}}$
that are orthogonal to each other and are a linear combination of
the old parameters ${\normalcolor \mathbf{O}}$, through the eigenvectors\begin{equation}
\mathbf{X}=\mathbf{EO}\,.\end{equation}
The $X_{i}$ are called the principal components of the experiment
and are ordered so that $X_{1}$ and $X_{n}$ are the best and worst
measured components respectively. In this construction, the eigenvalues
are the variances of the new parameters so that $X_{1}$ has the smallest
eigenvalue and $X_{n}$ has the largest one. The eigenvectors have
been normalised so that $\sum_{j}e_{j}^{2}=1$, where $e_{j}$s are
the elements of $E_{i}$. We list some properties of PCA below;

\begin{itemize}
\item The main point of PCA is to assess the degeneracies amongst the parameters
that are not resolved by the experiments, be they fundamental like
the cosmic variance or due to the noise and coverage of the experiment.
In our case, it will especially help us to see the correlation amongst
the bins of the primordial PS, and between the bins and the cosmological
parameters.
\item The eigenvalues obtained measure the performance of the experiment
--- a larger number of small eigenvalues means a better experiment.
Another measure of the performance of the experiments is to see how
they mix physically independent parameters such as, say, $n_{s}$,
the spectral index, and $\Omega_{b}$. This sort of mixture may be
improved by improving the experiment.%
\footnote{However, the so-called 'geometrical degeneracy' \citep{ZSS-GeometricalDeg,EB-GeometricalDeg}
will not be improved by improving the experiments; two models with
same primordial PS, the same matter content, and the same comoving
distance to the surface of last scattering produce identical CMB PS.%
}
\end{itemize}
All the above points may be summed up to conclude that in a perfect
setting we would expect a one-to-one relation between the old and
the new parameters. This means that we would see only one of the old
parameters to contribute completely to one of the new parameters,
with zero contribution from the other old parameters.

Note that the principal components obtained are not unique and depend
on th\textcolor{black}{e form} of the variables (e.g., whether we
use $\Omega_{b}$ or $\log\Omega_{b}$), \textcolor{black}{as well
as where they are evaluated. }

\subsection{Hermitian Square Root}

Another approach to remove the correlations between the uncertainties
is to use the Hermitian square root of the Fisher matrix as a linear
transformation on the parameter space \citep{bjk,Hamilton1,Hamilton2}\textcolor{black}{.}
This transformation matrix is obtained by\begin{equation}
\mathbf{F}^{1/2}=\mathbf{E^{T}}\mathbf{\Lambda}^{1/2}\mathbf{E}\,,\end{equation}
where, like before, $\mathbf{E}$ is the eigenvector matrix and $\mathbf{\Lambda}$
is a diagonal matrix containing the eigenvalues. It has the property
$\mathbf{F}=\mathbf{F}^{1/2}\mathbf{F}^{1/2}=(\mathbf{F}^{1/2})^{\mathbf{T}}\mathbf{F}^{1/2}$
and therefore the condition $(\mathbf{F}^{-1/2})\mathbf{F}(\mathbf{F}^{-1/2})=(\mathbf{F}^{-1/2})^{\mathbf{T}}\mathbf{F}(\mathbf{F}^{-1/2})=diag$
is satisfied. Unlike PCA, $\mathbf{F}^{1/2}$ does not give us an
orthogonal basis and instead, it can be thought of as 'window functions'
for the primordial PS. We define a window matrix by\begin{equation}
H_{nm}=\frac{(F^{1/2})_{nm}}{\sum_{n}(F^{1/2})_{nm}}\,,\label{eq:norm_Herm}\end{equation}
which satisfies the normalisation condition $\sum_{n}H_{nm}=1$. Hence
the windowed PS is defined as\begin{equation}
\tilde{P}_{m}=\sum_{n}H_{nm}P(k_{n})\,,\end{equation}
where $P(k_{n})$ is the original primordial PS. Note that this windowed
PS is not a physically motivated PS and it is just constructed for
a visual presentation and understanding of the underlying correlations.
However, in a perfect setting we would expect this windowed PS to
be equal to the primordial PS (i.e., with each window function comprising
a single bin).

We obtain this window matrix for the marginalised Fisher matrix of
the bins and hence it can only be applied to the Fisher matrices of
Planck and the combination of Planck and SDSS, which are invertible.

\section{Results}

\subsection{PCA}

First we show the results for the PCA. Principal components, the $X_{i}$s,
obtained for SDSS, Planck and their combination are shown as colour-coded
matrix plots; $X_{i}$s are shown from left to right with increasing
errors (which is equal to $1/\Lambda_{i}^{1/2}$, as the eigenvalues
are constructed for the Fisher matrix). Original parameters are shown
vertically starting with the bins on the bottom to $\theta_{i}$s
on the top. For the bins, the vertical width of the box is an indication
of the binwidth. Some $k$ values for the bins are shown on the figures.
We group the components depending on the $X_{i}$, showing those with
values greater than 0.4; those between $0.2$ and $0.4$; and those
below $0.2$.

\textcolor{black}{\underbar{SDSS}}

\begin{figure}
\includegraphics[width=1.0\columnwidth]{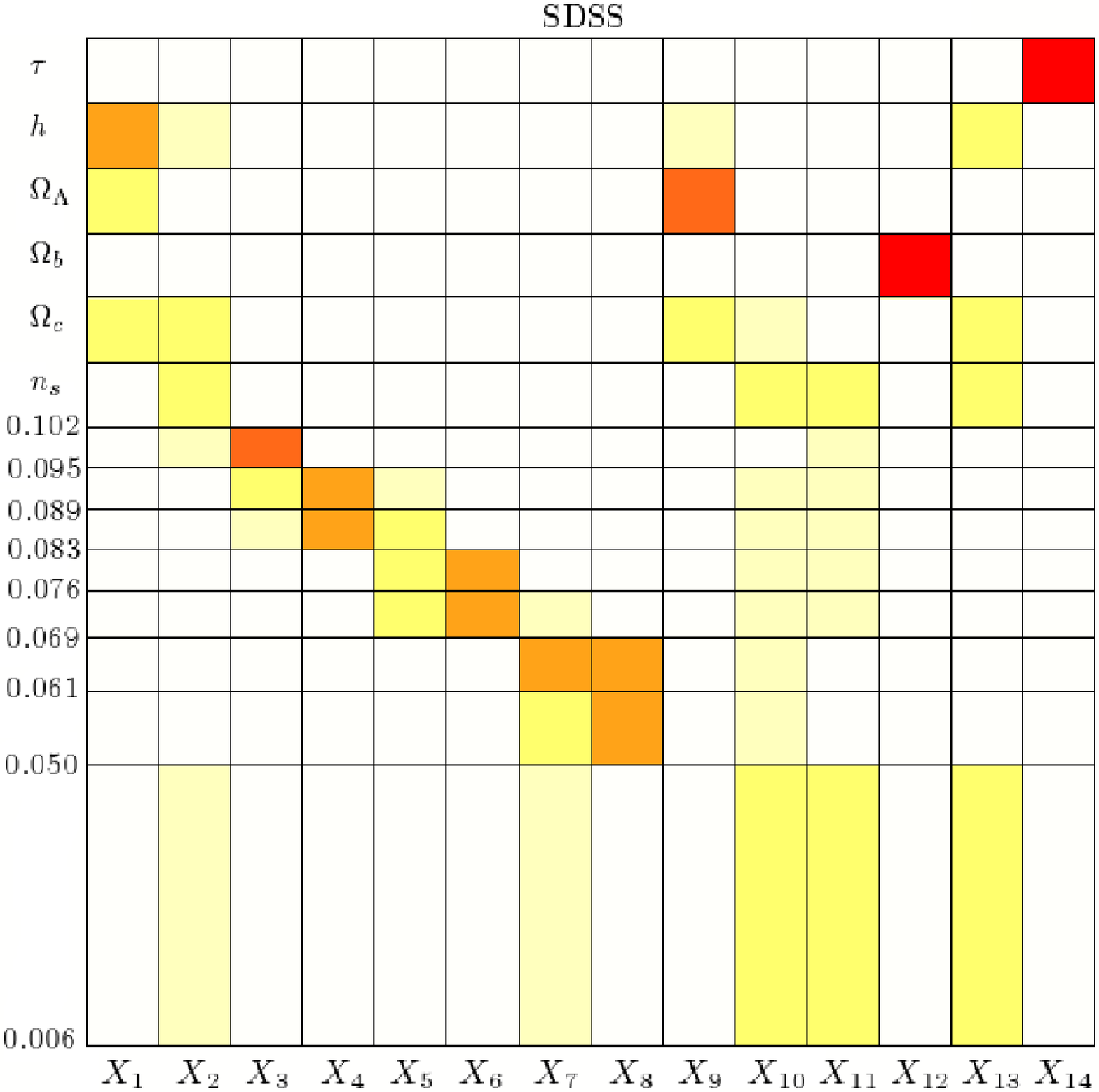}
\centering
\includegraphics[scale =0.35]{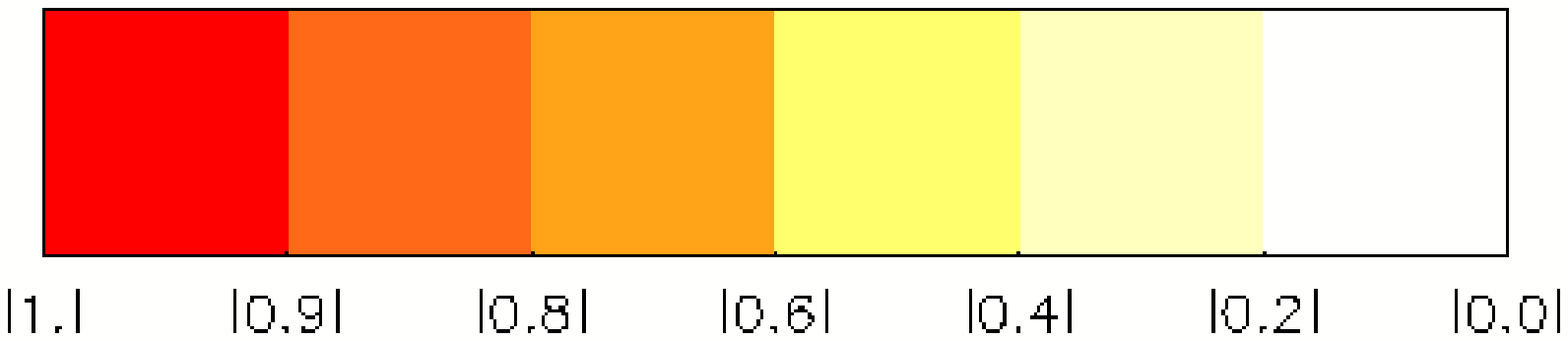}
\caption{{\small The principal components for SDSS with no
priors on $\theta_{i}$s. $X_{i}$ are shown from left to right with
increasing errors ($=1/\sqrt{\Lambda_{i}}$). Original parameters
are shown vertically starting with the bins on the bottom to $\theta_{i}$s
on the top. For the bins, the vertical width of the box is an indication
of the binwidth. We group the components depending on the $X_{i}$, showing those with
values greater than 0.4; those between $0.2$ and $0.4$; and those
below $0.2$. (refer to Figure \ref{fig:Colour-Plot} for the colour coding). The
last $6$ principal components can be ignored as they are not measured
--- refer to Table \ref{tab:Errors} and text for more details. At the bottome we show the colour plot indicating
different levels of contribution to the principal components.\label{fig:Set2-SDSS}}}
\end{figure}

The result is shown in Figure \ref{fig:Set2-SDSS}. There are a total
of $8$ bins that could be obtained to meet the $S/N$ criteria as
explained above. Together with the $6$ $\theta_{i}$s, we have a
total of $14$ original parameters and $14$ principal components,
$X_{i}$s. The last $6$ principal components are not measured well
(they have large/negative eigenvalues --- Table \ref{tab:Errors}).
This is because, as explained before, the SDSS Fisher matrix is not
a positive definite matrix; we have a total of $8$ data points and
this means only $8$ parameters (or $8$ different combinations of
the parameters, i.e. $X_{i}$s) can be measured.

\begin{figure}

\includegraphics[width=1.0\columnwidth]{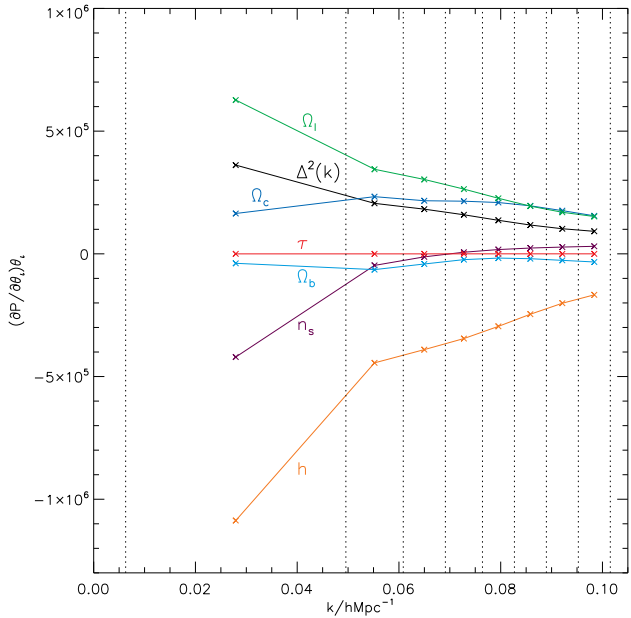}
\caption{{\small The derivative of galaxy PS with
respect to the primordial PS bins and $\theta_{i}$s, weighted by the
parameters values. This is exactly what goes in the Jacobian. It is
interesting to see in this $k$-range, much of the variation is on
large scales. Hence it is no surprise that SDSS measures small
scales better. \label{fig:dPdparams}}}

\end{figure}

\begin{figure}
\includegraphics[width=1.0\columnwidth]{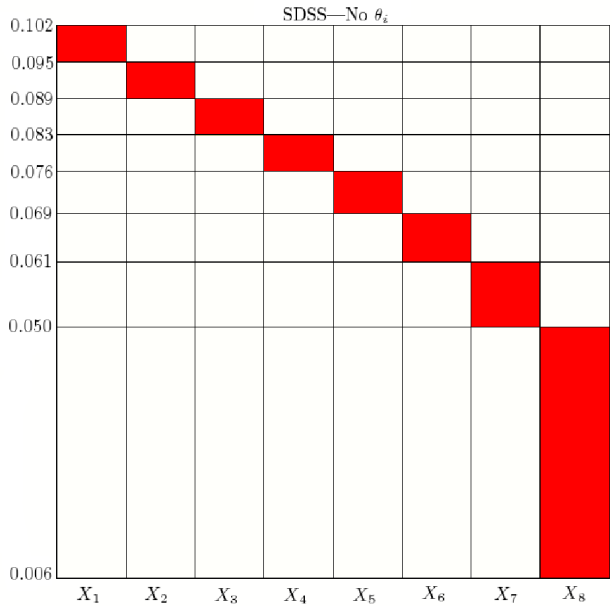}

\caption{{\small The principal components of SDSS for the primordial
PS bins only, assuming $\theta_{i}$s are known perfectly. No correlation
exists between the bins. Compare to Figures \ref{fig:Set2-SDSS} and
\ref{fig:Set4-SDSS}, and see how the lack of knowledge of the cosmological
parameters induce correlation between (neighbouring) bins. \label{fig:Set5-SDSS}}}

\end{figure}

The best measured principal component, $X_{1}$, has only the cosmological
parameters ($\theta_{i}$) contributing significantly, with $h$ being
dominant. The fact that there is more than one cosmological parameter
contributing to this principal component means that SDSS can only
measure a linear combination of them --- a degeneracy between these
parameters. $X_{2}$ measures a combination of the bins and $\theta_{i}$s.
Other principal components, $X_{3}$-$X_{8}$, measure the bins only,
with no contribution from $\theta_{i}$s at all, and the correlation
amongst the bins is between neighbouring ones only. Intuitively, you
would expect more correlation between the bins and $\theta_{i}$s.
Remember that the errors for the bins are related to the matter transfer
function --- equation \ref{eq:primfishmat}. Therefore, you would
expect that a change in $\theta_{i}$s would induce a change in the
matter transfer function and hence a correlation between bins and
$\theta_{i}$s. However, look at Figure \ref{fig:dPdparams} where
it is showing all the derivatives that goes in the Jacobian. The derivatives
with respect to $\theta_{i}$s seem to scale relatively close (apart
from $\tau$ where it had to be multiplied by $200$). However, the
derivative with respect to the primordial PS bins has to be rescaled
by $10^{-8}$ to fit in the same range as the rest of the derivatives.
This suggests that perhaps the changes in $\theta_{i}$s are not large
enough in this setting to have a significant effect on the matter
transfer function and therefore the correlation is not that significant
to show effects in the PCA. Note that the correlation between the
bins shows our limits to what we can learn about the primordial PS.
This correlation arises due to our lack of knowledge of the cosmological
parameters. If we knew the parameters perfectly, we would have what
is shown \textcolor{black}{in Figure }\ref{fig:Set5-SDSS}, which
is in fact the Fisher matrix itself. Generally it seems that SDSS
measures cosmological parameters better than the primordial PS and
within primordial PS bins, it measures small scales better than large
scales.

\begin{figure}
\includegraphics[width=1.0\columnwidth]{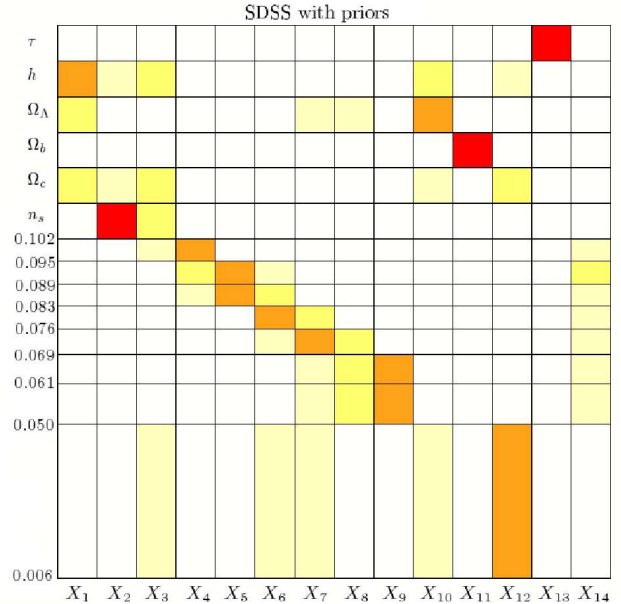}

\caption{{\small The principal components for SDSS with WMAP5
priors. Now all $X_{i}$s, apart from $X_{14}$, are measured well
--- Table \ref{tab:Errors}. Also, some of the degeneracies between
$\theta_{i}$s have been broken. \label{fig:Set4-SDSS}}}

\end{figure}

We also investigated what improvements we would see given better ---
realistic --- knowledge of the cosmological parameters. Hence, WMAP5
priors \citep{Dunkley-WMAP} were added to constrain the $\theta_{i}$s
in the Fisher matrix, by adding the inverse variance of each parameter
to the Fisher matrix, i.e. ignoring the correlations. The result is
shown in Figure \ref{fig:Set4-SDSS}. Some of the degeneracies between
the cosmological parameters have been broken. For example, $X_{2}$
now measures $n_{s}$ almost perfectly. Also, $\Omega_{b}$ and $\tau$
dominate completely in $X_{11}$ and $X_{12}$ respectively, with
no contribution from any other parameter. Note that the errors on
the principal components have reduced and now all $X_{i}$s, except
$X_{14}$, can be measured well --- Table \ref{tab:Errors}. This
is expected as WMAP5 does a good job measuring these cosmological
parameters. With respect to bins, it seems that adding priors and
improving constraints on cosmological parameters has only helped to
measure linear combinations of the bins better and has not been able
to break the degeneracy between them.

\underbar{Planck}\label{Planck:-For-Planck}

\begin{figure}
\includegraphics[width=1.0\columnwidth]{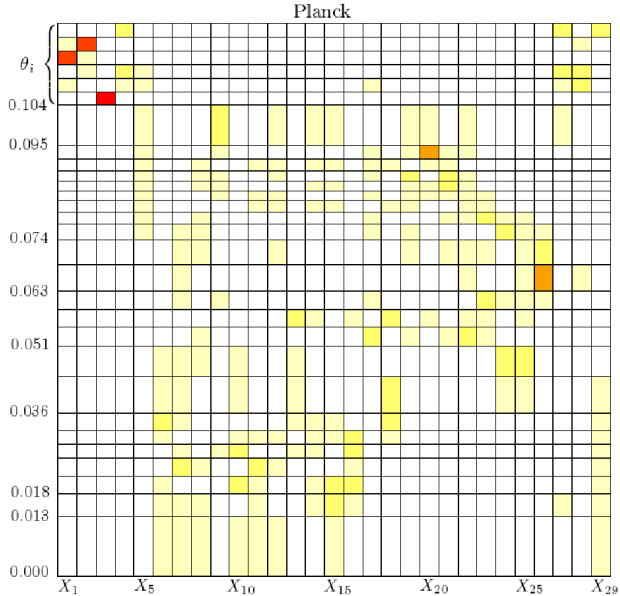}

\caption{{\small The principal components of Planck. Again
bins are shown on the bottom and cosmological parameters on the top,
ordered in the same way shown in previous figures. All principal components
seem to be measured well and better than the SDSS case. Although,
no particular scale seem to dominate strongly to any of the principal
components. \label{fig:Set2-Planck}}}

\end{figure}

\begin{figure}
\includegraphics[width=1.0\columnwidth]{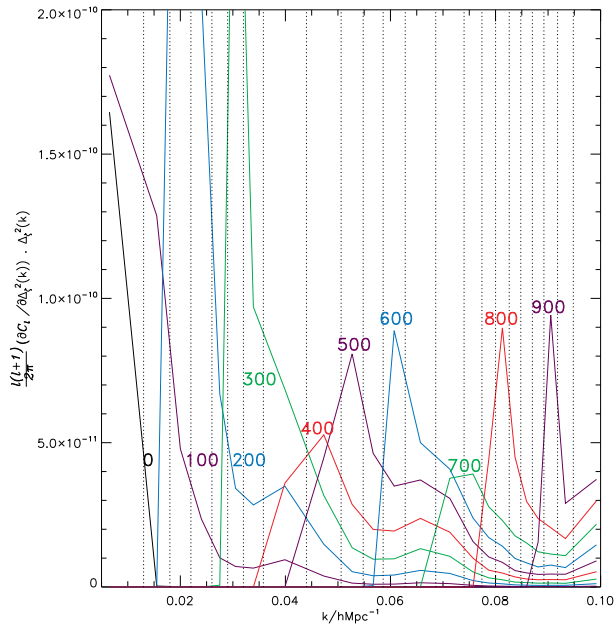}

\caption{{\small The derivative of radiation PS with respect
to the primordial PS bins, equation \ref{eq:Cl_pPS_derivative}, weighted
by the primordial PS. Note that the bin with $\ell=400$ dominating,
gets contributions from all $\ell$s from $100$ to $500$. This makes
the correlation between the bins on all scales possible.\label{fig:dCldpPS}}}

\end{figure}

For Planck there are a total of $23$ bins and this, with the $6$
$\theta_{i}$s, means we have $29$ principal components, shown in
Figure \ref{fig:Set2-Planck}. They all seem to be measured well and
better than SDSS --- Table \ref{tab:Errors}. The reflection of the
\textcolor{black}{acoustic peaks of $C_{\ell}$s on the bin sizes
can clearly be seen; the ones corresponding to the peaks are measured
with a better} resolution. To see this, look at equation\begin{equation}
F_{\alpha\beta}^{\prime}=\sum_{\ell}\left[\delta_{\ell\ell^{\prime}}F_{\ell\ell^{\prime}}(C_{\ell})^{2}\right]\left[\frac{\partial C_{\ell}}{\partial\theta_{\alpha}}\Delta_{\zeta}^{2}(k_{\alpha})\right]\left[\frac{\partial C_{\ell^{\prime}}}{\partial\theta_{\beta}}\Delta_{\zeta}^{2}(k_{\beta})\right]\,,\end{equation}
which is $(S/N)_{\alpha\beta}^{2}$. This is equation \ref{eq:CMB FM} weighted by the signals. First bracket
can be ignored as it is almost a constant due to the relation $F\propto C_{l}^{-2}$
--- equation \ref{eq:FM_CMB}. The primordial PS, $\Delta_{\zeta}^{2}(k_{i})$,
in the second and third brackets can also be ignored as it is a constant.
Therefore we are left with \begin{equation}
F_{\alpha\beta}^{\prime}\propto\sum_{\ell}\frac{\partial C_{\ell}}{\partial\theta_{\alpha}}\frac{\partial C_{\ell^{\prime}}}{\partial\theta_{\beta}}\,.\end{equation}
For the bins the derivative in this equation is the radiation transfer
function as shown in equation \ref{eq:Cl_pPS_derivative}. The summation
over $\ell$ then gives the oscillatory feature seen in $k$ space
--- see Figure \ref{fig:dCldpPS} to see the pictorial version of
this.

Just like SDSS, Planck seems to measure the cosmological parameters
better than the primordial PS and overall does a better job than SDSS,
giving smaller errors and less correlation between them. This is not
surprising as we already know Planck does a good job measuring the
cosmological parameters; it measures $\Omega_{\Lambda}$, $h$ and
$n_{s}$ very well, with only slight correlation with other cosmological
parameters. Note that $n_{s}$ is measured almost perfectly with no
correlation with $\theta_{i}$s (or the bins, as expected).

The rest of principal components contain the highly-correlated bins
only, with no particular large contribution from any of them. Intuitively
one might expect the correlation to be between neighbouring bins only.
The reason for the longer-range correlation lies in the form of the
radiation transfer function; for each $\ell$, this transfer function
spans a $k$-range around $k$$=\ell/d_{A}$, where $d_{A}$ is the
angular-diameter distance to the last-scattering surface. This is
due to the projection of a $3$D Universe onto a $2$D sphere around
us. Equation \ref{eq:Cl_pPS_derivative} shows what exactly contributes
to the Jacobian for the Fisher matrix analysis. For each $\ell$,
this derivative integrates the radiation transfer function over the
$k$-range of the bins. This would be reflected as correlation between
neighbouring bins. However, remember that in the Fisher matrix analysis
the $\ell$s get summed over (equation \ref{eq:CMB FM}) and this
now makes correlation between all bins possible; Figure \ref{fig:dCldpPS}
shows a pictorial version of equation \ref{eq:Cl_pPS_derivative},
weighted by the primordial PS. Note how each $\ell$ spans a range
of $k$. The summation over all $\ell$s means that, for example,$\:$the
bin with $\ell=400$ dominating has contributions from all $\ell$s
from $100$ to $500$, with each spanning a different range of $k$.
This induces correlation between bins of all scales.

\begin{figure}
\includegraphics[width=1.0\columnwidth]{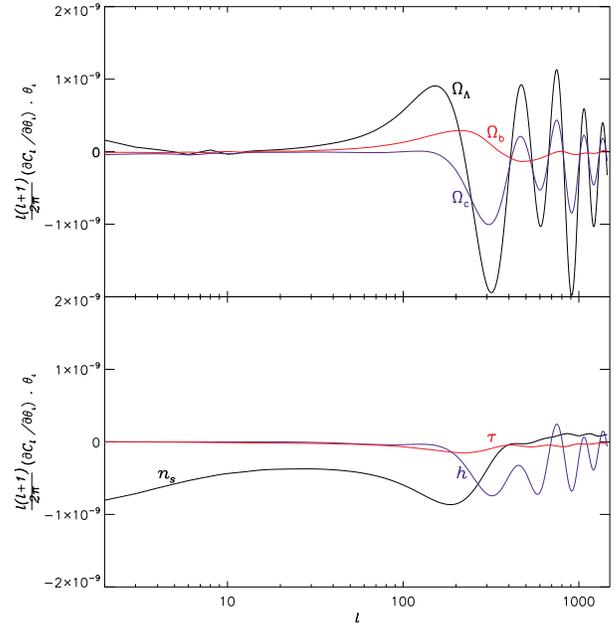}

\caption{{\small The derivative of radiation PS with respect
to $\theta_{i}$s. Note how different parameters dominate on different
scales. For example, $n_{s}$ dominates on large scales and $\Omega_{b}$
dominates on small scales. If$\:$Planck can only measure a linear
combination of $n_{s}$ and $\Omega_{b}$, the degeneracy between
these parameters could induce a degeneracy between large and small
scale bins! \label{fig:dCldparams1}}}

\end{figure}

This sort of correlation between small and large scales might even
be worse when there is a degeneracy between the measured cosmological
parameters. For example, consider an experiment that could only measure
a linear combination of $n_{s}$ and $\Omega_{b}$, where $n_{s}$
is dominant on large scales and $\Omega_{b}$ is dominant on small
scales --- Figure \ref{fig:dCldparams1}. The degeneracy between these
parameters could induce a degeneracy between large and small scale
bins.

\begin{figure}
\includegraphics[width=1.0\columnwidth]{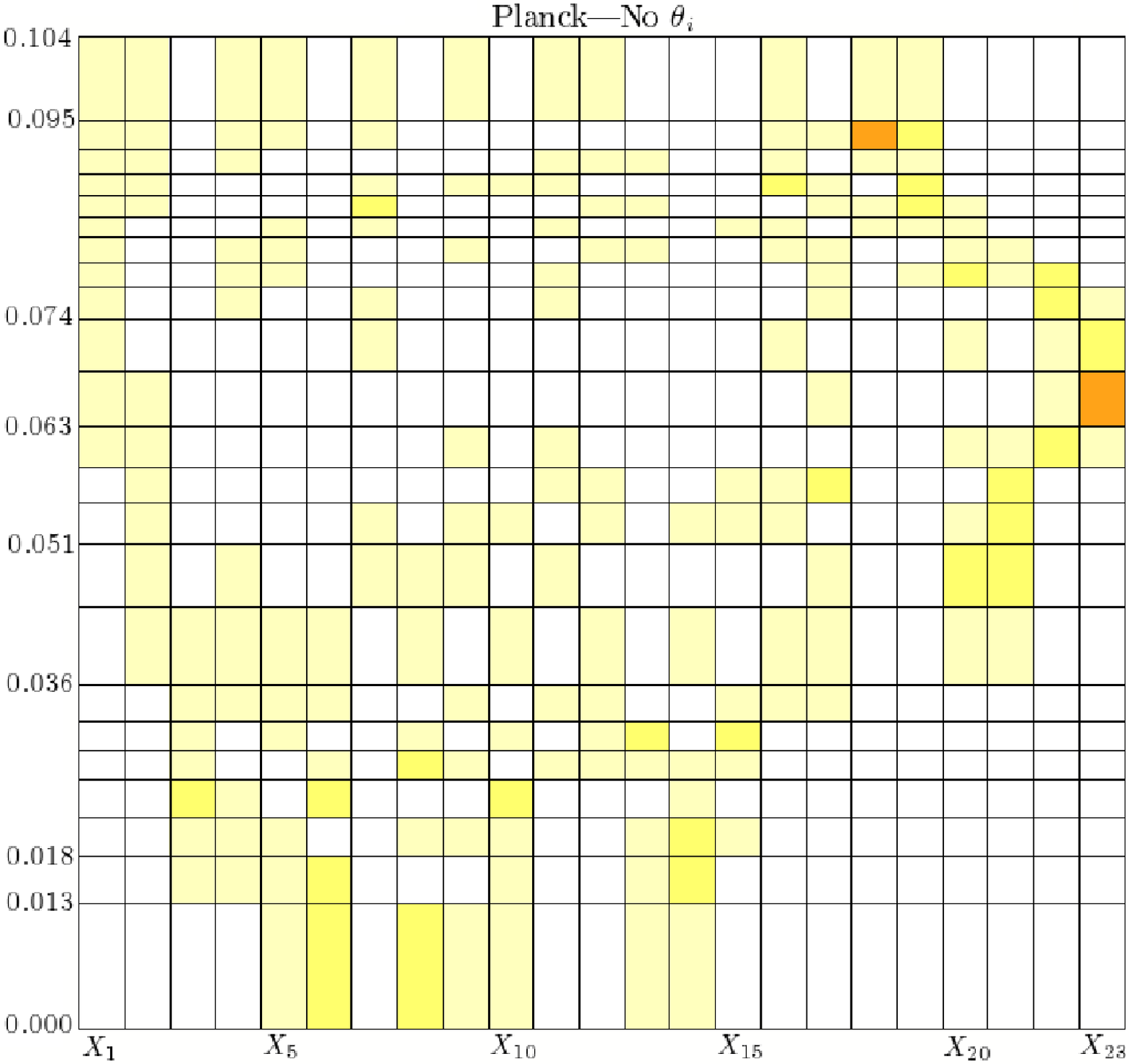}

\caption{{\small The principal components of Planck for the
primordial PS bins only, assuming $\theta_{i}$s are known perfectly.
Compare to Figure \ref{fig:Set2-Planck}. It seems like lack of knowledge
of cosmological parameters does not have much of an effect in measuring
the primordial PS bins in this case. This could be due to the fact
that Planck measures $\theta_{i}$s very well.\label{fig:Set5-Planck}}}

\end{figure}

We also investigated if the lack of prior knowledge of $\theta_{i}$s
induces extra correlation between the bins, as in the SDSS case. Figure
\ref{fig:Set5-Planck} shows the principal components for the bins
with no $\theta_{i}$s --- i.e. assuming cosmological parameters are
known perfectly. \textcolor{black}{Since Planck's measurements of
the parameters will be much better than even those from WMAP (inclusion
of which was able to remove the correlations for SDSS), we might expect
little change. Indeed, not much is changed}. The only improvement
is on the range of$\:$ errors, which now span a smaller range ---
Table \ref{tab:Errors}. Note, however, that the smallest error for
this set is still larger than the smallest error for the set including
$\theta_{i}$s. This is because $\theta_{i}$s are generally measured
better than the primordial PS bins and hence they lower the errors.
Instead, comparing the largest errors of both sets shows the improvements.
Despite the smaller errors for this set, not much is improved in terms
of correlation between the bins.

\begin{figure}
\includegraphics[width=1.0\columnwidth]{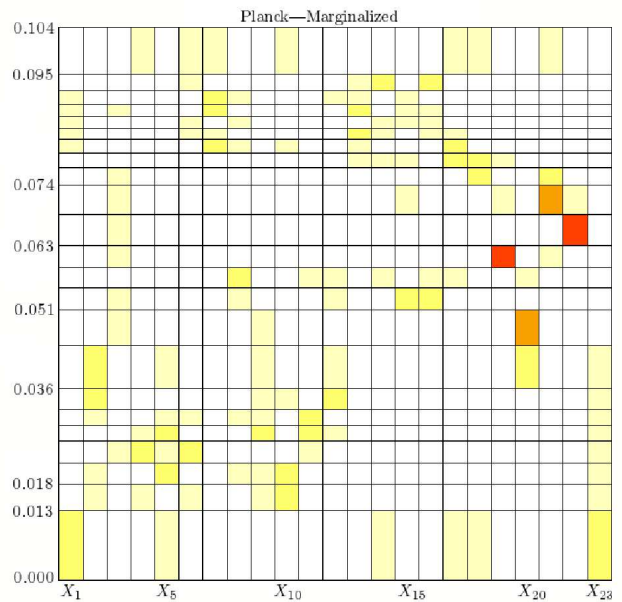}

\caption{{\small The principal components of the 'marginalised'
Fisher matrix of Planck.\label{fig:marg-Planck}}}

\end{figure}

We also want to consider the bins on their own. Hence, we consider
the correlation between the bins for the marginalised Fisher matrix
of bins (that is, marginalised over the other cosmological parameters,
$\theta_{i}$\textcolor{black}{s}). This is obtained by inverting
the parent Fisher matrix to get a covariance matrix, which holds the
marginalised errors for all the parameters. Take the sub-block of
this matrix which holds the errors for the bins and invert this to
get a marginalised Fisher matrix and diagonalise this matrix. The
principal components for this Fisher matrix are shown in Figure \ref{fig:marg-Planck}.
The first thing to note is that bins contribute more significantly
to some of the principal components. In particular there are some
mid-scale bins which seem to be measured well. For example, look at
$X_{19}$ and $X_{22}$; they seem to have uncorrelated some mid-scale
bins from the rest of the bins.

\textcolor{black}{Another interesting result is that very large and
very small scales never really dominate in the principal components
with large errors. They only contribute to them at levels of $\lesssim0.01$.
Remember that $X_{i}$s with large errors carry the most correlation
and therefore the fact that mid-scale bins do not contribute to these
principal components means that they are measured quite well.}

To sum up, it seems like Planck will largely decorrelate the primordial
PS from the $\theta_{i}$s (and therefore the transfer function) but
cannot exactly uncorrelate the bins themselves.

\underbar{Planck \& SDSS}

\begin{figure}
\includegraphics[width=1.0\columnwidth]{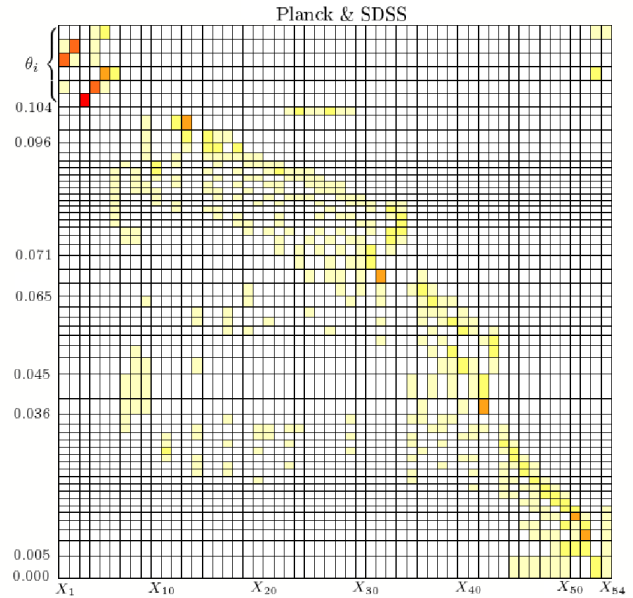}

\caption{{\small The principal components of Planck \& SDSS.
Clearly, resolution of the primordial PS has improved. Also, an almost
diagonal trend can be seen now, showing small scales are measured
better than the large scales. There is also less correlation between
$\theta_{i}$s. \label{fig:Set2-Planck&SDSS}}}

\end{figure}

The results are shown in Figure \ref{fig:Set2-Planck&SDSS}. Combining
surveys has clearly helped to improve the resolution of the primordial
PS. Now there are a total of $48$ bins in the same $k$-range. Again
the cosmological parameters are measured better than the primordial
PS and there is also less correlation between the cosmological parameters
compared to the previous cases. There is also less correlation between
the bins themselves. Both features of SDSS and Planck can clearly
be seen here. For example, acoustic oscillations in the $C_{\ell}$s
still influence the bin sizes and resolution of the primordial PS.
It also seems like small scales are measured better than large scales,
which is a feature seen in the SDSS case.

\begin{figure}
\includegraphics[width=1.0\columnwidth]{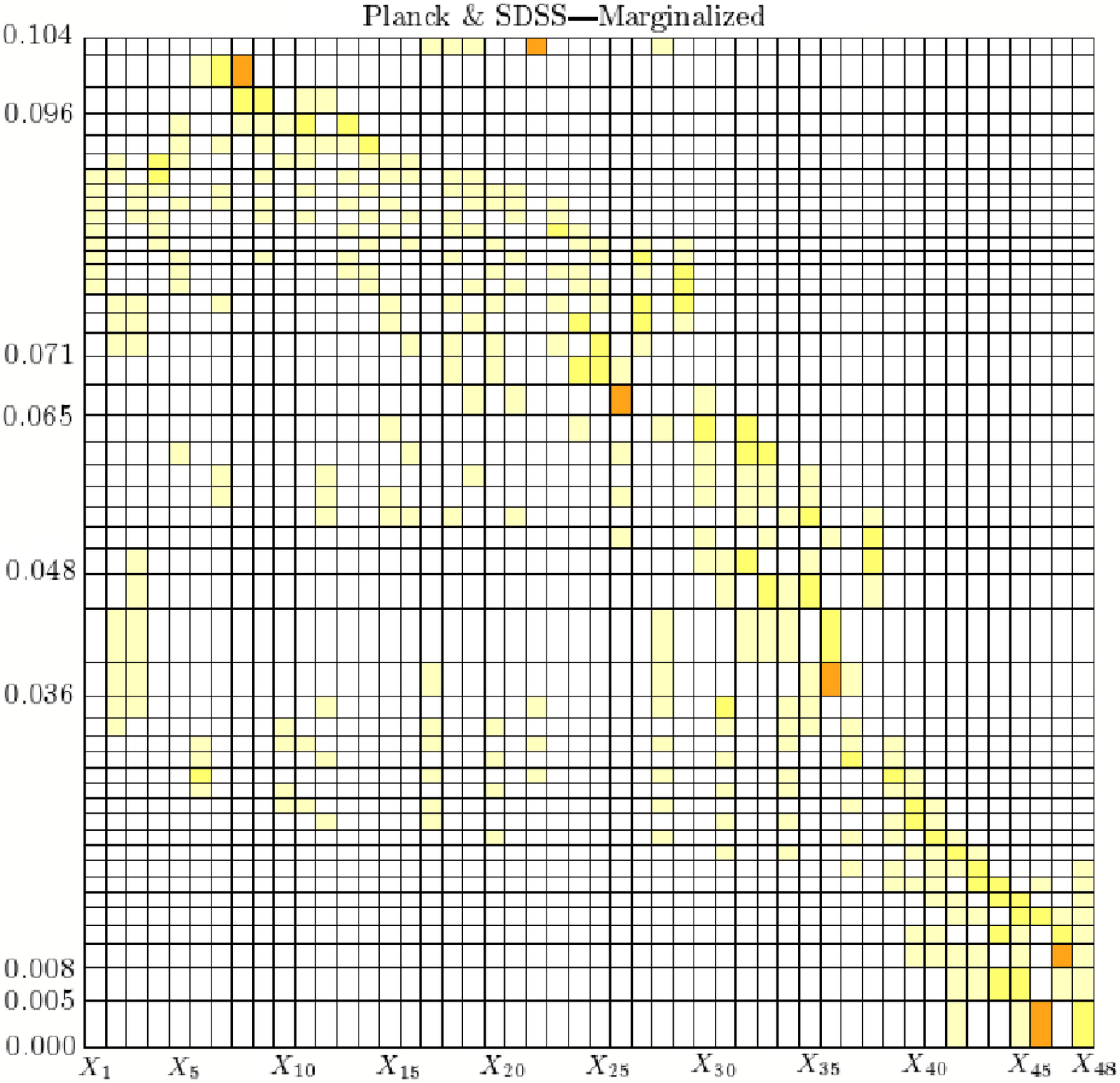}

\caption{{\small The principal components of the 'mariginalised'
Fisher matrix of Planck \& SDSS.\label{fig:marg-Planck&SDSS}}}

\end{figure}

Figure \ref{fig:marg-Planck&SDSS} shows the results for the marginalised
Fisher matrix of the bins for SDSS and Planck combined. Compare to
Figure \ref{fig:Set2-Planck&SDSS}; not much change can be seen.

\subsection{Hermitian Square Root of Fisher Matrix}

\begin{figure}
\includegraphics[width=1.0\columnwidth]{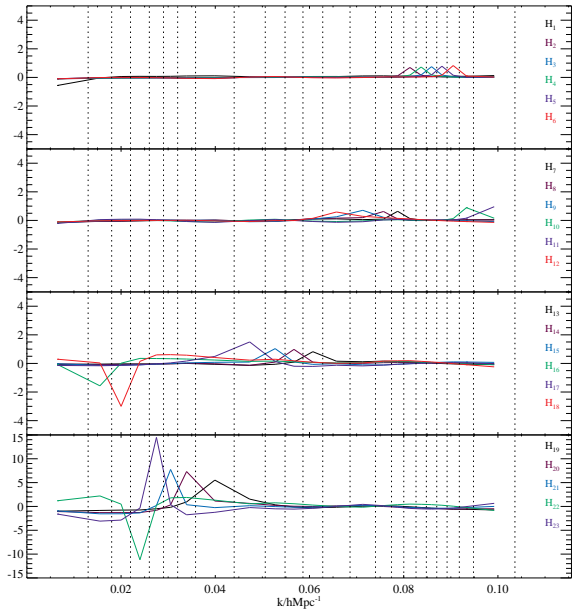}

\caption{{\footnotesize \label{fig:FMCl-Herm}}{\small The row
vectors of $\underline{H}$ for the marginalised Fisher matrix of
Planck. These vectors are ordered with increasing errors, so that
$H_{1}$ is the best and $H_{23}$ is the worst measured vector. This,
unlike the principal components, shows that correlation is only between
}\textit{\small neighbouring}{\small{} bins and, that bins on large
scales are measured better than the ones on small scales. }}

\end{figure}

\begin{figure}
\includegraphics[width=1.0\columnwidth]{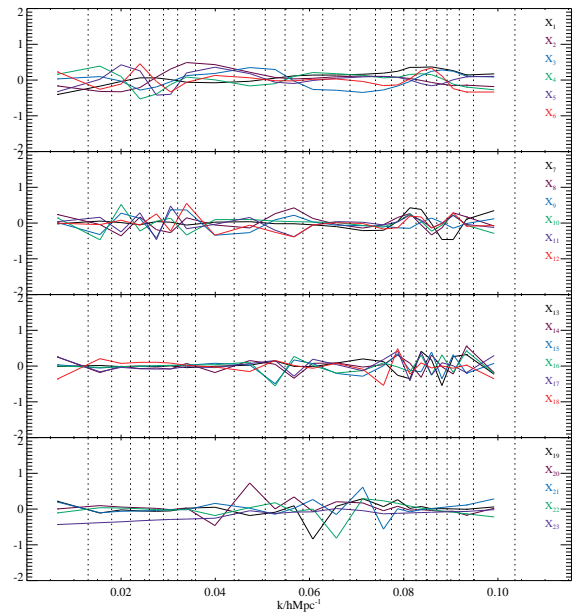}

\caption{\label{fig:eigvCl-HermFormat}{\small This is Figure \ref{fig:marg-Planck}
plotted in this way for easier comparison with Figure \ref{fig:FMCl-Herm}.}}

\end{figure}

Figure \ref{fig:FMCl-Herm} shows the window functions for Planck
derived from the Hermitian square root decorrelation. Note that only
the magnitude of the components of $H_{m}$s are important and not
their sign. However, it is worth mentioning that for the non-marginalised
Fisher matrix (both for Planck and its combination with SDSS), these
window functions have only positive values. Therefore, it is the lack
of knowledge of the cosmological parameters (and the induced correlation
between the bins) that introduces non-physical negative values into
the window functions. The window functions, $H_{m}$s, are plotted
in the order of increasing errors, so that $H_{1}$ is the best measured
and $H_{23}$ the worst measured vector, respectively. Here, small
scales seem to be measured best and large scales measured worst, contributing
to $H_{m}$s with the lowest and highest errors respectively. It seems
like Planck has not been able to decorrelate the bins completely and
some correlations between \textit{neighbouring} bins can be seen.
In addition, bins in the range of $k\sim0.02-0.04h\textnormal{Mpc}^{-1}$
have a large contribution to their $H_{m}$s, compared to the other
bins. Compare this to Figure \ref{fig:eigvCl-HermFormat}, where we
diagonalised the marginalised Fisher matrix through its eigenvectors
(This is exactly Figure \ref{fig:marg-Planck} plotted in this format
for easier comparison). In the PCA case, the correlations seem not
to be only between neighbouring bins, but between bins of all scales,
which is not seen in this case! Also, the compactness seen here (i.e.
more of a window-type feature) is not seen in the PCA case; there
is no particular scale that contributes significantly to the principal
components.

\begin{figure}
\includegraphics[width=1.0\columnwidth]{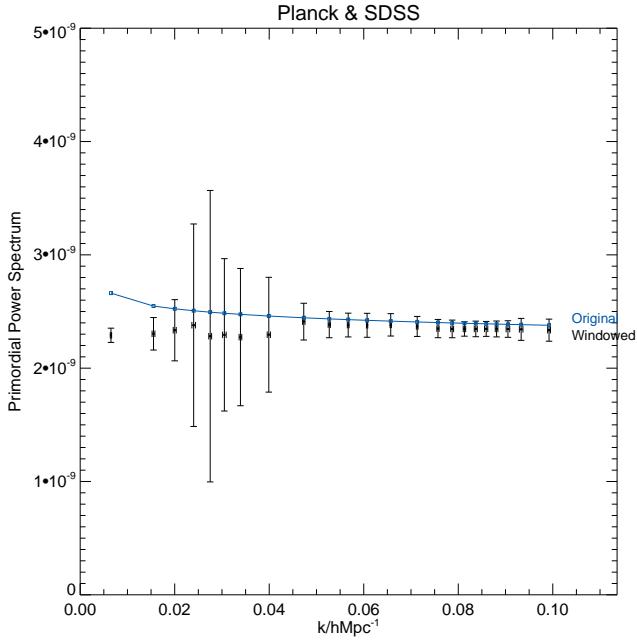}

\caption{{\footnotesize \label{fig:FMCl-HermPS}}{\small The
windowed PS obtained from Planck.}}

\end{figure}

Figure \ref{fig:FMCl-HermPS} shows the windowed PS for Planck. It
is plotted so that each $\tilde{P}_{m}$ is placed at the $k_{n}$
from which it receives the largest contribution. The vertical errors
bars shown are $\Delta_{\zeta}^{2}(k_{i})(\mathbf{H}\mathbf{F}^{-1}\mathbf{H}^{T})$,
where $\Delta_{\zeta}^{2}(k_{i})$ is the amplitude of the primordial
PS in the bins and $(\mathbf{H}\mathbf{F}^{-1}\mathbf{H}^{T})$ is
the errors propagated through the $H_{m}$ distribution. The horizontal
error bars are the half-width at half-maximum in each direction of
the main peak of each $H_{m}$. The original primordial PS is plotted
for comparison. $\tilde{P}_{m}$ seems to be at a lower level than
the unwindowed primordial PS. Remember that $\tilde{P}_{m}$ is not
a physical PS \textit{per se}. However, the observed differences from
the original PS arise due to the induced correlations between the
bins. In a perfect setting, where there are no correlations between
bins, you do expect $\tilde{P}_{m}=\Delta_{\zeta}^{2}(k)$ to be true.
Note that the main feature of this plot is that vertical errors, unlike
those for the original primordial PS, are \textit{not} correlated.
The correlation between the errors has been transferred to overlaps
between the window functions --- as shown in Figure \ref{fig:FMCl-Herm}.

\begin{figure}
\includegraphics[width=1.0\columnwidth]{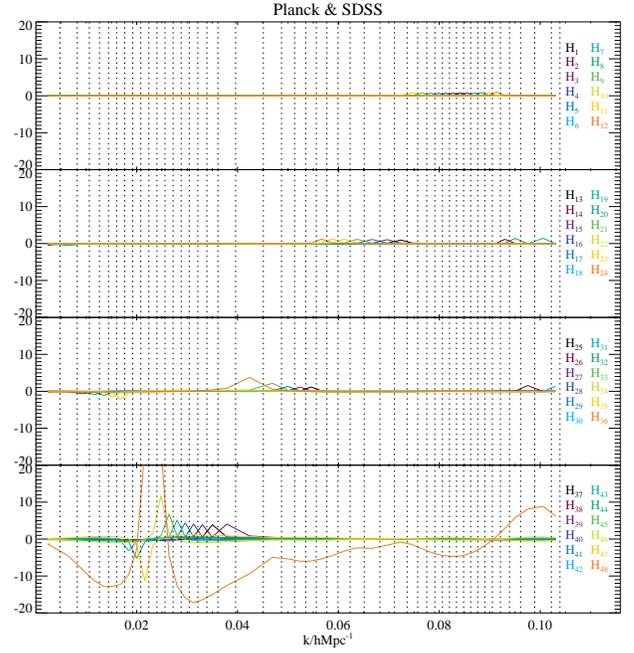}

\caption{{\footnotesize \label{fig:FMPCl-Herm}}{\small The
row vectors of $\underline{H}$ for the marginalised Fisher matrix
of Planck \& SDSS. Like before, they are ordered with increasing errors.
The correlation between neighbouring bins still exists but to a lesser
extent. Also, note that the bins are narrower here so that correlation
between neighbouring bins still means a correlation within a narrower
$k$-range.}}

\end{figure}

\begin{figure}
\includegraphics[width=1.0\columnwidth]{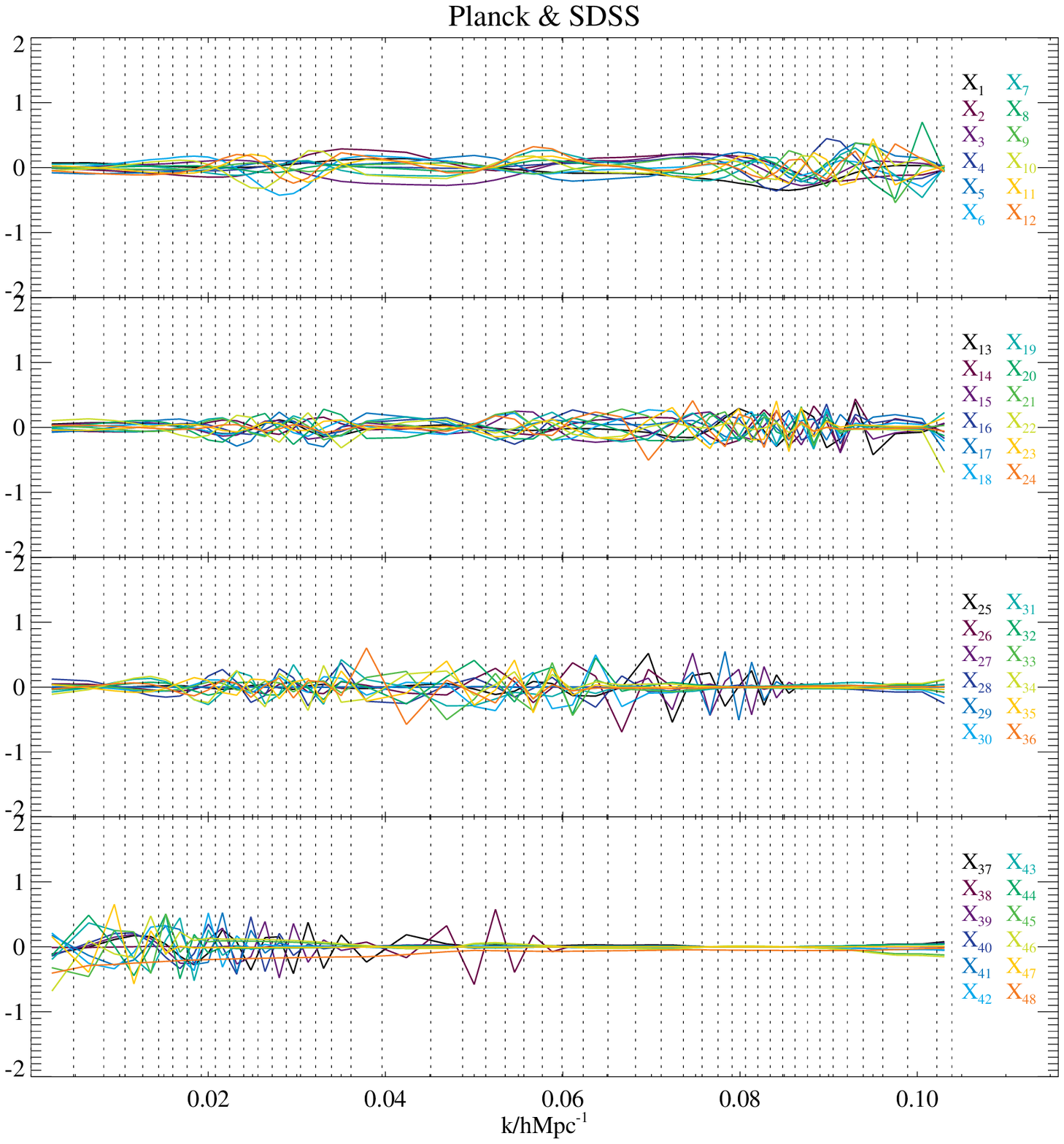}

\caption{\label{fig:eigvPCl-HermFormat}{\small This is Figure \ref{fig:marg-Planck&SDSS}
plotted in this way for easier comparison with Figure \ref{fig:FMPCl-Herm}.}}

\end{figure}

\begin{figure}
\includegraphics[width=1.0\columnwidth]{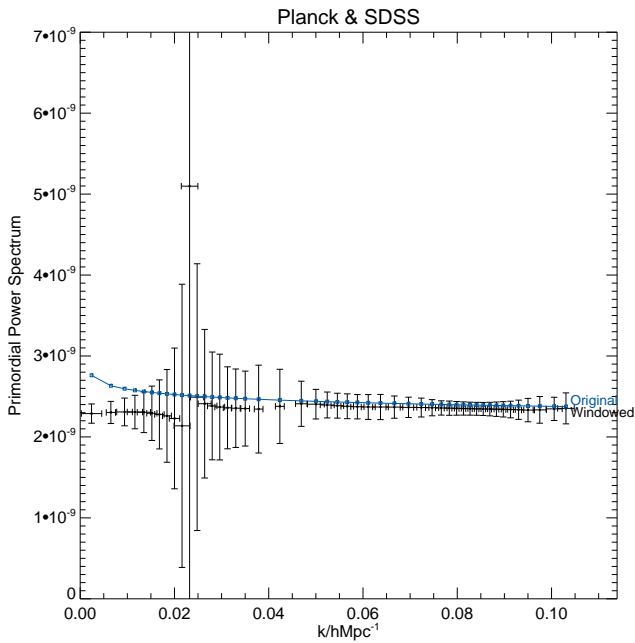}
\caption{{\footnotesize \label{fig:FMPCl-HermPS}}{\small The
windowed PS obtained from combination of Planck \& SDSS.}}
\end{figure}

Figures \ref{fig:FMPCl-Herm} and \ref{fig:FMPCl-HermPS} show the
same set of results for combination of Planck and SDSS. Again large
scales are contributing to $H_{m}$s with the largest errors. There
seems to be less correlation between neighbouring bins compared to
the Planck case. Also, note that bins in this case are narrower and
therefore correlation between neighbouring bins still means correlation
between a narrower range of $k$. Compare Figure \ref{fig:FMPCl-Herm}
to Figure \ref{fig:eigvPCl-HermFormat} (same as Figure \ref{fig:marg-Planck&SDSS}).
Again, there is less compactness in the PCA case, however more than
what is seen for Planck on its own. Figure \ref{fig:FMPCl-Herm} indcates
that bins in the vicinity of $k\sim0.02-0.025h\textnormal{M}\textnormal{pc}^{-1}$
seem to contribute very strongly to their $H_{m}$s compared to other
bins, in particular the last window function, $H_{48}$. This effect
gets carried on to $\tilde{P}_{m}$, with $\tilde{P}_{11}$ having
a very large amplitude --- Figure \ref{fig:FMPCl-HermPS}.

\section{Conclusions}

The primordial PS holds precious information about the physics of
the early Universe and constraining it has been one of the key goals
of the modern cosmology. However, the induced degeneracy between the
cosmological parameters of the matter/radiation transfer functions
and the primordial PS limit our ability to recover the primordial
PS completely, even from a perfect survey, especially in the case
of CMB surveys \citep{ho}. Different surveys \textcolor{black}{probe
different scales with different accuracies} and might not be able
to constrain the primordial PS to a desired resolution on their own.
However, put together, they make significant improvements. In this
paper we have investigated these limits/improvements for Planck and
SDSS. For this purpose, we have assumed a non-parametric function
of the primordial PS and have constructed a parameter space containing
a set of \textit{carefully chosen} bins of the primordial PS along
with a set of cosmological parameters. We constructed a Fisher matrix
for this parameter space for the two different surveys separately
and combined. By diagonalising these Fisher matrices, via two different
methods of eigenvector decomposition (PCA) and the Hermitian square
root, we have investigated the induced correlation between the primordial
PS bins and the cosmological parameters.

In the PCA case, we came to conclude that SDSS and Planck together
measure the cosmological parameters to a better extent, and even break
the degeneracy between them. They can increase the resolution of the
primordial PS by about twice as much and can also condense the correlation
between bins to be only amongst neighbouring ones. On the whole it
seems like they can constrain small scales better than large scales.

By the use of Hermitian square root of the Fisher matrix we managed
to divert the correlation amongst the marginalised errors of the bins
to the correlation between the bins themselves. In this case, combination
of SDSS and Planck helped to decrease the level of correlation between
neighbouring bins, but also, because it has helped to increase the
resolution of the bins, correlation between neighbouring bins means
correlation between a smaller range of $k$.

Clearly adding the two surveys have helped to constrain the primordial
PS to a better degree. Obviously, further surveys such as Ly-$\alpha$
(e.g. SDSS Ly$\alpha$F PS), weak lensing (e.g. Euclid), peculiar
velocity (e.g. Cluster Imaging Experiment (CIX)), etc. can help even
more to measure the primordial PS.

\acknowledgements{We especially thank Carlo Contaldi for his great suggestions. We
thank Dmitri Novikov, George Bendo, Daniel Mortlock and Gavin Nicholson
for their great help. This work was supproted by STFC.}

\clearpage{}

\begin{table}[h]
\centering
\caption{\label{tab:Errors}{\small Errors for different sets for SDSS, Planck
and combination of Planck and SDSS.}}
\begin{tabular}{|>{\centering}p{1.1in}>{\centering}p{0.36in}>{\centering}p{0.36in}>{\centering}p{0.36in}>{\centering}p{0.36in}>{\centering}p{0.36in}>{\centering}p{0.36in}>{\centering}p{0.36in}>{\centering}p{0.36in}>{\centering}p{0.36in}|} 
\hline
\noalign{\vskip0.02in}
{\small SDSS} & {\small $X_{1}$} & {\small $X_{2}$} & {\small $X_{3}$} & {\small $X_{5}$} & {\small $X_{7}$} & {\small $X_{9}$} & {\small $X_{11}$} & {\small $X_{13}$} & {\small $X_{14}$}\tabularnewline
\hline
\noalign{\vskip0.02in}
{\small No priors} & \textcolor{black}{\small $0.0038$} & {\small $0.0160$} & {\small $0.0287$} & \textcolor{black}{\small $0.0328$} & {\small $0.0339$} & {\small $7\textnormal{E}5$} & {\small NaN} & {\small NaN} & \textcolor{black}{\small NaN}\tabularnewline
{\small No $\theta_{i}$} & \textcolor{black}{\small $0.0282$} & {\small $0.0299$} & {\small $0.0317$} & \textcolor{black}{\small $0.0340$} & {\small $0.0359$} & {\small ---} & {\small ---} & {\small ---} & \textcolor{black}{\small ---}\tabularnewline
{\small WMAP5 priors} & \textcolor{black}{\small $0.0038$} & {\small $0.0123$} & {\small $0.0236$} & \textcolor{black}{\small $0.0308$} & {\small $0.0340$} & {\small $0.0357$} & {\small $0.0709$} & {\small $0.02571$} & \textcolor{black}{\small $23.38$}\tabularnewline
\hline
\noalign{\vskip0.2in}
\hline
\noalign{\vskip0.025in}
{\small Planck} & {\small $X_{1}$} & {\small $X_{2}$} & {\small $X_{3}$} & {\small $X_{5}$} & {\small $X_{10}$} & {\small $X_{15}$} & {\small $X_{20}$} & {\small $X_{25}$} & {\small $X_{28/29}$}\tabularnewline
\hline
\noalign{\vskip0.02in}
{\small PCA-No priors} & \textcolor{black}{\small $0.0004$} & {\small $0.0022$} & {\small $0.0035$} & \textcolor{black}{\small $0.0152$} & \textcolor{black}{\small $0.0295$} & \textcolor{black}{\small $0.0402$} & \textcolor{black}{\small $0.0515$} & \textcolor{black}{\small $0.0700$} & \textcolor{black}{\small $0.4953$}\tabularnewline
{\small PCA-No $\theta_{i}$s} & \textcolor{black}{\small $0.0110$} & {\small $0.0149$} & {\small $0.0181$} & \textcolor{black}{\small $0.0224$} & \textcolor{black}{\small $0.0333$} & \textcolor{black}{\small $0.0422$} & \textcolor{black}{\small $0.0572$} & \textcolor{black}{\small ---} & \textcolor{black}{\small ---}\tabularnewline
{\small PCA-Margin.} & {\small $0.0204$} & {\small $0.0210$} & {\small $0.0255$} & \textcolor{black}{\small $0.0313$} & \textcolor{black}{\small $0.0410$} & {\small $0.0553$} & {\small $0.0890$} & {\small ---} & {\small ---}\tabularnewline
Hermitian Sqrt & {\small $0.0236$} & {\small $0.0261$} & {\small $0.0268$} & {\small $0.0283$} & {\small $0.0386$} & {\small $0.0487$} & {\small $0.2525$} & {\small ---} & {\small ---}\tabularnewline
\hline
\noalign{\vskip0.2in}
\hline
\noalign{\vskip0.02in}
{\small Planck \& SDSS} & {\small $X_{1}$} & {\small $X_{2}$} & {\small $X_{3}$} & {\small $X_{10}$} & {\small $X_{20}$} & {\small $X_{30}$} & {\small $X_{40}$} & {\small $X_{50}$} & {\small $X_{54}$}\tabularnewline
\hline
\noalign{\vskip0.02in}
{\small PCA-No priors} & \textcolor{black}{\small $0.0004$} & {\small $0.0020$} & {\small $0.0035$} & \textcolor{black}{\small $0.0348$} & \textcolor{black}{\small $0.0463$} & \textcolor{black}{\small $0.0548$} & \textcolor{black}{\small $0.0642$} & \textcolor{black}{\small $0.1132$} & \textcolor{black}{\small $0.5261$}\tabularnewline
{\small PCA-No $\theta_{i}$s} & \textcolor{black}{\small $0.0561$} & {\small $0.0568$} & {\small $0.0575$} & \textcolor{black}{\small $0.0624$} & \textcolor{black}{\small $0.0729$} & \textcolor{black}{\small $0.0891$} & \textcolor{black}{\small $0.1003$} & \textcolor{black}{\small ---} & {\small ---}\tabularnewline
{\small PCA-Margin.} & {\small $0.0254$} & {\small $0.0289$} & {\small $0.0304$} & {\small $0.0425$} & {\small $0.0523$} & {\small $0.0598$} & {\small $0.0823$} & {\small ---} & {\small ---}\tabularnewline
Hermitian Sqrt & {\small $0.0323$} & {\small $0.0325$} & {\small $0.0327$} & {\small $0.0374$} & {\small $0.0578$} & {\small $0.0762$} & {\small $0.2102$} & {\small ---} & {\small ---}\tabularnewline
\hline
\end{tabular}
\end{table}

\bibliographystyle{apj}
\bibliography{biblio}

\end{document}